\documentclass[12pt]{article}
\usepackage{epsfig}

\newskip\humongous \humongous=0pt plus 1000pt minus 1000pt

\newif\ifdtup

\def\pr#1{#1^\prime}

\def\beq{\begin{equation}}
\def\eeq{\end{equation}}

\def\beqn{\begin{eqnarray}}
\def\eeqn{\end{eqnarray}}
\relax

\catcode`\@=11

\@addtoreset{equation}{section}
\def\theequation{\thesection.\arabic{equation}}

\def\@normalsize{\@setsize\normalsize{15pt}\xiipt\@xiipt
\abovedisplayskip 14pt plus3pt minus3pt%
\belowdisplayskip \abovedisplayskip
\abovedisplayshortskip \z@ plus3pt%
\belowdisplayshortskip 7pt plus3.5pt minus0pt}

\def\small{\@setsize\small{13.6pt}\xipt\@xipt
\abovedisplayskip 13pt plus3pt minus3pt%
\belowdisplayskip \abovedisplayskip
\abovedisplayshortskip \z@ plus3pt%
\belowdisplayshortskip 7pt plus3.5pt minus0pt
\def\@listi{\parsep 4.5pt plus 2pt minus 1pt
     \itemsep \parsep
     \topsep 9pt plus 3pt minus 3pt}}

\@twosidetrue

\relax

\catcode`@=12

\catcode`\@=11

\def\section{\@startsection{section}{1}{\z@}{3.5ex plus 1ex minus
   .2ex}{2.3ex plus .2ex}{\large\bf}}

\def\thesection{\arabic{section}}

\def\appendix{\setcounter{section}{0}
 \def\thesection{\Alph{section}}
 \def\theequation{\Alph{section}.\arabic{equation}}
 \def\section{\@startsection{section}{1}{\z@}{3.5ex plus 1ex minus
   .2ex}{2.3ex plus .2ex}{\large\bf}}
 \def\subsection{\@startsection{section}{2}{\z@}{3.25ex plus 1ex minus
   .2ex}{1.5ex plus .2ex}{\large\bf}}}

\def\ps@headings{\def\@oddfoot{}\def\@evenfoot{}
\def\@oddhead{\hbox{}\hfill
 \makebox[.5\textwidth]{\raggedright\ignorespaces --\thepage{}--
 \hfill {}}}  
\def\@evenhead{\@oddhead}
\def\subsectionmark##1{\markboth{##1}{}}
}

\ps@headings

\catcode`\@=12

\def\figcap{\section*{Figure Captions\markboth
 {FIGURECAPTIONS}{FIGURECAPTIONS}}\list
 {Fig. \arabic{enumi}:\hfill}{\settowidth\labelwidth{Fig. 999:}
 \leftmargin\labelwidth
 \advance\leftmargin\labelsep\usecounter{enumi}}}
 \relax
\def\tablecap{\section*{Table Captions\markboth
 {TABLECAPTIONS}{TABLECAPTIONS}}\list
 {Table \arabic{enumi}:\hfill}{\settowidth\labelwidth{Table 999:}
 \leftmargin\labelwidth
 \advance\leftmargin\labelsep\usecounter{enumi}}}
 \relax
\def\reflist{\section*{References\markboth
 {REFLIST}{REFLIST}}\list
 {[\arabic{enumi}]\hfill}{\settowidth\labelwidth{[999]}
 \leftmargin\labelwidth
 \advance\leftmargin\labelsep\usecounter{enumi}}}
 \relax

\catcode`\@=11

\relax

\catcode `@ 11
\def\biblabel#1{\if@filesw\immediate
\write\@auxout{\string\bibcite{#1}{\the\value{\@listctr }}}\fi}
\catcode `@ 12

\newcommand{\ccaption}[2]{
  \begin{center}
    \parbox{0.85\textwidth}{
      \caption[#1]{\small\it {#2}}}
  \end{center}    }
\def    \be             {\begin{equation}}
\def    \ee             {\end{equation}}
\def    \ba             {\begin{eqnarray}}
\def    \ea             {\end{eqnarray}}

\def    \=              {\;=\;}
\def    \frac           #1#2{{#1 \over #2}}

\def \as   {\ifmmode \alpha_s \else $\alpha_s$ \fi}

\def\b0{b_0}

\def \mt   {\ifmmode m_{\rm t} \else $m_{\rm t}$ \fi}

\def \to   {\mbox{$\rightarrow$}}
\newcommand     \MSB            {\ifmmode {\overline{\rm MS}} \else
                                 $\overline{\rm MS}$\fi}

\newcommand\hepph[1]{{\tt hep-ph/#1}}

\def\figura#1#2#3
{\begin{figure}[hb]
  \begin{center}
  \leavevmode\protect\epsfxsize=#1\protect\epsffile{#2.eps}
  \protect\caption{#3}
  \label{#2}
  \end{center}
  \end{figure}
}
\newlength{\Largfig}
\def\({\left(}
\def\){\right)}

\def\s{\sigma}
\def\Qb{\overline{Q}}
\def\asb{{}\ifmmode \bar{\alpha}_s \else $\bar{\alpha}_s$\fi}
\def\Dh{\hat{D}}
\def\t{\,\log\frac{\mu^2}{\mu_0^2}\,}

\relax
\def\pl#1#2#3{{\it Phys. Lett. }{\bf #1}\ (19#2)\ #3}

\def\pr#1#2#3{{\it Phys. Rev. }{\bf #1}\ (19#2)\ #3}
\def\np#1#2#3{{\it Nucl. Phys. }{\bf #1}\ (19#2)\ #3}

\relax

\begin{document}
\begin{titlepage}
\nopagebreak
{\flushright{
        \begin{minipage}{4cm}
        CERN-TH/97-209  \hfill \\
        IFUM 588/FT \hfill \\
        hep-ph/9709358\hfill \\
        \end{minipage}        }

}
\vfill
\begin{center}
{\LARGE \bf \sc
 \baselineskip 0.9cm
On the Fragmentation Function for\newline
Heavy Quarks in $e^+e^-$ collisions

 }
\vskip 2cm
{\bf Paolo NASON\footnote{On leave of absence from INFN, Milan, Italy},}
\\
\vskip 0.1cm
{CERN, TH Division, Geneva, Switzerland} \\
\vskip .5cm
{\bf Carlo OLEARI}
\\
\vskip .1cm
{Dipartimento di Fisica, Universit\`a di Milano and INFN, Milan, Italy}
\end{center}
\nopagebreak
\vfill
\begin{abstract}
  We use a recent ${\cal O}(\as^2)$ calculation of the differential cross
  section for the production of heavy quarks in $e^+e^-$ annihilation to 
  compute a few moments of the heavy quark single-inclusive production
  cross section. We verify that, contrary to some recent claims,
  the leading and next-to-leading logarithmic terms in this cross section
  are correctly given by the standard NLO fragmentation function formalism
  for heavy quark production.
\end{abstract}
\vskip 1cm
CERN-TH/97-209 \hfill \\
\today \hfill
\vfill
\end{titlepage}

\section{Introduction}

It is known that inclusive heavy quark production is a calculable process
in perturbative QCD, since the heavy quark mass acts as a cut-off for the
final state collinear singularities.
Thus, the process
\beq
  e^+ e^- \,\to\, Z/\gamma\,\to\, Q+ X\;,
\eeq
where $Q$ is the heavy quark and $X$ is anything else, is calculable.
Its cross section can be expressed as a power expansion in the strong
coupling constant
\beq
\label{eq:sigma}
   \frac{d\sigma}{dx}(x,E,m)
= \sum_{n=0}^{\infty} a^{(n)}(x,E,m,\mu)\, \asb^n(\mu)\;,
\eeq
where $E$ is the centre-of-mass energy, $m$ is the mass of the heavy quark,
$\mu$ is the renormalization scale, and
\beq
 \asb(\mu) =  \frac{\as(\mu)}{2\pi}\;.
\eeq
As usual we define
\beq
   x = \frac{2\, p\cdot q}{q^2}\,,
\eeq
where $q$ and $p$ are the four-momenta of the intermediate virtual 
boson and of the final heavy quark $Q$.
The cross section~(\ref{eq:sigma}), normalized to the total cross section,
is sometimes referred to as the heavy quark fragmentation function in $e^+e^-$
annihilation.
When $E/m$ is not too large, the truncation of eq.~(\ref{eq:sigma})
at some fixed order in the coupling can be used to compute the cross section.
On the other hand, if $E\gg m$, the $n^{\rm th}$ order coefficient
of the expansion will in general contain up to $n$ powers of $\log \(E/m\)$,
thereby spoiling the convergence of the expansion. These large logarithms
can be resummed, according to the method described in ref.~\cite{MeleNason}.
First of all, since $\log\(E/m\)$ is large, 
one is entitled to neglect the terms
that are suppressed by powers of $m/E$.
One then observes that, in this limit, the inclusive heavy quark cross
section must satisfy the factorization theorem formula
\beq\label{eq:factorization}
   \frac{d\s}{dx}(x,E,m) =
\sum_i \int_x^1 \frac{dz}{z}\, 
 \frac{d\hat\s_i}{dz}\(z,E,\mu\) \,
  \hat{D}_i\(\frac{x}{z},\mu,m\)\;,
\eeq
where $d\hat\s_i(z,E,\mu)/dz$ are the \MSB-subtracted partonic
cross sections for producing the parton $i$, and $\hat{D}_i(x,\mu,m)$
are the \MSB\ fragmentation functions for the parton $i$ into the
heavy quark $Q$.  In order for eq.~(\ref{eq:factorization}) to hold,
it is essential that one uses a renormalization scheme where the heavy
flavour is treated as a light one, like the pure \MSB\ scheme. Thus
$d\hat\s_i(z,E,\mu)/dz$ has a perturbative expansion in terms of $\as$
with $n_f$ flavours, where $n_f$ includes the heavy one.
The scale $\mu$ is the factorization and renormalization scale. It
should be chosen of the order of $E$, in order to avoid the appearance
of large logarithms of $E/\mu$ in the partonic cross section.
The \MSB\ fragmentation functions  $\hat{D}_i$ obey the Altarelli-Parisi
evolution equations:
\beq
\label{eq:AP}
  \frac{d \hat D_i}{d\log\mu^2} (x,\mu,m) = \sum_j\int^1_x \frac{dz}{z}\,
   P_{ij}\(\frac{x}{z},\asb(\mu)\) \;\hat{D}_j(z,\mu,m)\;.
\eeq
The Altarelli-Parisi splitting functions $P_{ij}$ have the perturbative
expansion
\beq
\label{eq:APfunctions}
    P_{ij}\Bigl(x,\asb(\mu)\Bigr) = \asb(\mu) P^{(0)}_{ij}(x) 
    + \asb^2(\mu) P^{(1)}_{ij}(x) + {\cal O}(\asb^3)\;,
\eeq
where $P_{ij}^{(0)}$ are given in ref.~\cite{AltarelliParisi}
and $P_{ij}^{(1)}$ have been computed in
refs.~\cite{CurciFurmanskiPetronzio}-\cite{Konishi}.
The only missing ingredients for the calculation of the inclusive cross
section are the initial conditions for the \MSB\ fragmentation functions.
These were obtained at the NLO level in ref.~\cite{MeleNason} by matching
the ${\cal O}(\asb)$ direct calculation of the process (i.e. 
formula~(\ref{eq:sigma})) with the expansion of 
formula~(\ref{eq:factorization})
at order $\asb$. They have the form
\beqn
\hat{D}_Q(x,\mu_0,m)&=&\delta(1-x)+\asb(\mu_0)\,
 d^{(1)}_Q(x,\mu_0,m)+{\cal O}(\asb^2)
\nonumber\\ \label{eq:ini}
\hat{D}_g(x,\mu_0,m)&=&\asb(\mu_0)\, d^{(1)}_g(x,\mu_0,m)+{\cal O}(\asb^2)\;,
\eeqn
all the other components being of order $\asb^2$.
Thus, in order to compute the NLO resummed expansion, one takes the
initial conditions eqs.~(\ref{eq:ini}), at a value of $\mu_0$ of order $m$,
evolves them at the scale $\mu$ (taken to be of order $E$), and then
applies formula~(\ref{eq:factorization}), using a NLO expression for
the partonic cross section
\beq\label{eq:sigmahat}
 \frac{d\hat\s_i}{dx}(x,E,\mu) =\hat{a}^{(0)}_i(x) + \hat{a}^{(1)}_i(x,E,\mu)
   \,\asb(\mu) + {\cal O}(\asb^2)\;.
\eeq
For example, if the parton $i$ is the heavy quark itself, one gets
\beq\label{eq:sigmahatQ}
 \frac{d\hat\s_Q}{dx}(x,E,\mu) =   \delta(1-x) + \hat{a}_Q^{(1)}(x,E,\mu)
   \,\asb(\mu) + {\cal O}(\asb^2) \;,
\eeq
where we have  normalized the cross section to 1 at zeroth order in 
the strong coupling constant.

The procedure outlined above guarantees that all terms of the form
$(\asb L)^n$ (leading order) and $\asb (\asb L)^n$ (next-to-leading order),
where $L$ is the large logarithm,
are included correctly in the resummed formula.
Observe that, at the NLO level, the scale that appears in $\asb$ in
eqs.~(\ref{eq:ini}) and~(\ref{eq:sigmahat}) could be changed by factors
of order 1, since this amounts to a correction of order $\asb^2$.
However, one cannot set $\mu_0=\mu$ in eqs.~(\ref{eq:ini})
(or $\mu=m$ in formula~(\ref{eq:sigmahat})), since this amounts to
a correction of order $\asb^2 L$, and thus it would spoil the
validity of the resummation formula at the NLO level.

There is essentially no room for these considerations to fail.
They are a consequence of the factorization theorem for fragmentation
functions, which is quite well established 
\cite{CutVertices,CurciFurmanskiPetronzio}. 

The validity of this procedure
has however been questioned by Kniehl et al.\ in ref.~\cite{Spira}.
In their procedure,
the heavy quark short distance cross section is replaced by
\beq\label{eq:sigmahatprime}
 \frac{d\hat\s^\prime_Q}{dx}(x,E,\mu) = \delta(1-x) +
 \asb(E)\,\left[ \hat{a}^{(1)}_Q(x,E,\mu) + d^{(1)}_Q(x,\mu_0,m) \right]
   + {\cal O}(\asb^2)\;,
\eeq
and the initial condition by
\beq\label{eq:iniprime}
\hat{D}^\prime_Q(x)=\delta(1-x)\;,
\eeq
which is to be evolved from the scale $\mu_0$ to the scale $\mu$
using the NLO \MSB\ evolution equations. This procedure differs
at the NLO level from the standard procedure advocated in
ref.~\cite{MeleNason}. The difference starts to show up in the
terms of order $\asb^2 L$.
Recently, we have completed a calculation of heavy 
quark inclusive production at order $\asb^2$~\cite{NasonOleari}.
Using this calculation,
we are in a position to verify explicitly the approach of
ref.~\cite{MeleNason}, and thereby dismiss the approach of ref.~\cite{Spira}.
In the next section we describe the procedure we followed in detail.

\section{Calculation}
Instead of dealing with the realistic case of $Z/\gamma$ decay,
we perform the calculation for a hypothetical vector boson $V$ that
couples only to the heavy quark with vectorial coupling.

We introduce the following notation for the Mellin transform:
\beq
   f(N) \equiv \int_0^1 dx \, x^{N-1} f(x)\;.
\eeq
We adopt the convention that, when $N$ appears instead of $x$ as the
argument of a function, we are actually referring to the Mellin transform
of the function. This notation is somewhat improper, but it should
not generate confusion in the following, since we will be working
only with Mellin transforms. The Mellin transform of the factorization 
formula~(\ref{eq:factorization}) is given by
\beq\label{eq:sigmaN}
\sigma(N,E,m) =\sum_i \hat\sigma_i(N,E,\mu) \; \hat{D}_i(N,\mu,m)\;,
\eeq
where
\beq
\sigma(N,E,m) = \int_0^1 dx \, x^{N-1} \frac{d\s}{dx}(x,E,m)\;,
\eeq
and a similar one for $\hat{\s}_i(N,E,\mu)$, 
and the Mellin transform of the Altarelli-Parisi
 evolution equation~(\ref{eq:AP}) is
\beq
\label{eq:APN}
 \frac{d\Dh_i(N,\mu,m)}{d\log\mu^2} = \sum_j 
 \asb(\mu) \left[ P_{ij}^{(0)}(N) + P_{ij}^{(1)}(N)\,
     \asb(\mu) +   {\cal O}(\asb^2)\right] \Dh_j(N,\mu,m)\;.
\eeq
We want to obtain an expression for $\sigma(N,E,m)$ valid at the second
order in $\asb$.
Thus, we need the solution of eq.~(\ref{eq:APN}), with initial condition
at $\mu=\mu_0$, accurate at order $\asb^2$. This is easily done by rewriting
eq.~(\ref{eq:APN}) as an integral equation
\beqn
&& \Dh_i(N,\mu,m) = \Dh_i(N,\mu_0,m)
 \nonumber \\ \label{eq:integAPN0} &&+
 \sum_j \int_{\mu_0}^\mu d\log{\mu^\prime}^2\;
 \asb(\mu^\prime) \left[ P_{ij}^{(0)}(N) + P_{ij}^{(1)}(N)\,
     \asb(\mu^\prime) \right]  \Dh_j(N,\mu^\prime,m)
\;.
\eeqn
The terms proportional to $\asb^2$ can be evaluated at any scale
($\mu$ or $\mu_0$),
the difference being of order $\asb^3$. Factors involving a single
power of $\asb$ can instead be expressed in terms of $\asb(\mu_0)$
using the renormalization group equation
\beqn
\asb(\mu^\prime)&=&\asb(\mu_0)-2\pi\,b_0\,\asb^2(\mu_0)\,
\log\frac{{\mu^\prime}^2}{\mu_0^2}+{\cal O}(\asb^3(\mu_0))
\cr
 b_0 &=& \frac{11C_A - 4\, n_f\,T_F}{12\pi}\,, 
\eeqn
where $n_f$ is the number of flavours including the heavy one.
Equation~(\ref{eq:integAPN0}) then becomes
\beqn
 \Dh_i(N,\mu,m) &=& \Dh_i(N,\mu_0,m)
+ \sum_j \int_{\mu_0}^\mu d\log{\mu^\prime}^2\;
 \asb(\mu_0) P_{ij}^{(0)}(N)
 \Dh_j(N,\mu^\prime,m)
 \nonumber \\  &+&
 \sum_j
 \asb^2(\mu_0) P_{ij}^{(1)}(N)
 \Dh_j(N,\mu_0,m)\;\log\frac{\mu^2}{\mu_0^2}
 \nonumber \\ \label{eq:integAPN1} &-&
 2\pi\,b_0\,\sum_j
 \asb^2(\mu_0) P_{ij}^{(0)}(N)
 \Dh_j(N,\mu_0,m)\;\frac{1}{2}\log^2\frac{\mu^2}{\mu_0^2}
\;.
\eeqn
We now need to express $\Dh_j(N,\mu^\prime,m)$ on the right-hand side
of the above equation as a function of the initial condition,
with an accuracy of order $\asb$. This is simply done by iterating
the above equation once, keeping only the first two terms on the 
right-hand side. Our final result is then
\beqn
 \Dh_i(N,\mu,m) &=& \Dh_i(N,\mu_0,m)
+ \sum_j
 \asb(\mu_0) P_{ij}^{(0)}(N)
 \Dh_j(N,\mu_0,m)\;\log\frac{\mu^2}{\mu_0^2}
 \nonumber \\ &+&
 \sum_{kj}
 \asb^2(\mu_0) P_{ik}^{(0)}(N)\;P_{kj}^{(0)}(N)
 \Dh_j(N,\mu_0,m)\;\frac{1}{2}\log^2\frac{\mu^2}{\mu_0^2}
 \nonumber \\  &+&
 \sum_j
 \asb^2(\mu_0) P_{ij}^{(1)}(N)
 \Dh_j(N,\mu_0,m)\;\log\frac{\mu^2}{\mu_0^2}
 \nonumber \\ \label{eq:integAPN2} &-&
 2\pi\,b_0\,\sum_j
 \asb^2(\mu_0) P_{ij}^{(0)}(N)
 \Dh_j(N,\mu_0,m)\;\frac{1}{2}\log^2\frac{\mu^2}{\mu_0^2}
\;.
\eeqn
Since the initial condition is
\beq
\Dh_j(N,\mu_0,m)=\delta_{jQ}+\asb(\mu_0)\,d^{(1)}_j(N,\mu_0,m)
   +{\cal O}(\asb^2(\mu_0))\,,
\eeq
eq.~(\ref{eq:integAPN2}) becomes, with the required accuracy:
\beqn
 \Dh_i(N,\mu,m) &=& \delta_{iQ}+\asb(\mu_0)\,d^{(1)}_i(N,\mu_0,m)
+  \asb(\mu_0) P_{iQ}^{(0)}(N)
 \;\log\frac{\mu^2}{\mu_0^2}
 \nonumber \\ &+&
 \sum_j
 \asb^2(\mu_0) P_{ij}^{(0)}(N)
 d^{(1)}_j(N,\mu_0,m)\;\log\frac{\mu^2}{\mu_0^2}
 \nonumber \\ &+&
 \sum_{k}
 \asb^2(\mu_0) P_{ik}^{(0)}(N)\;P_{kQ}^{(0)}(N)
 \;\frac{1}{2}\log^2\frac{\mu^2}{\mu_0^2}
 \nonumber \\  &+&
 \asb^2(\mu_0) P_{iQ}^{(1)}(N)\,
 \log\frac{\mu^2}{\mu_0^2}
 \nonumber \\ \label{eq:integAPN3} &-&
 2\pi\,b_0\,
 \asb^2(\mu_0) P_{iQ}^{(0)}(N)
 \;\frac{1}{2}\log^2\frac{\mu^2}{\mu_0^2}
\;.
\eeqn
Re-expressing $\asb(\mu_0)$ in terms of $\mu$, we get
\beqn
 \Dh_i(N,\mu,m) &=& \delta_{iQ}+\asb(\mu)\,d^{(1)}_i(N,\mu_0,m)
+ 2\pi\,b_0\,\asb^2(\mu)\,d^{(1)}_i(N,\mu_0,m)\;\log\frac{\mu^2}{\mu_0^2}
 \nonumber \\ &+&
  \asb(\mu) P_{iQ}^{(0)}(N)
 \;\log\frac{\mu^2}{\mu_0^2}
+ \sum_j
 \asb^2(\mu) P_{ij}^{(0)}(N)
 d^{(1)}_j(N,\mu_0,m)\;\log\frac{\mu^2}{\mu_0^2}
 \nonumber \\ &+&
 \sum_{k}
 \asb^2(\mu) P_{ik}^{(0)}(N)\;P_{kQ}^{(0)}(N)
 \;\frac{1}{2}\log^2\frac{\mu^2}{\mu_0^2}
+ \asb^2(\mu) P_{iQ}^{(1)}(N)\,
 \log\frac{\mu^2}{\mu_0^2}
 \nonumber \\ \label{eq:integAPN} &+&
 \pi\,b_0\,
 \asb^2(\mu) P_{iQ}^{(0)}(N)
 \;\log^2\frac{\mu^2}{\mu_0^2}
\;.
\eeqn
The partonic cross sections are given by
\beq\label{eq:partN}
\hat\sigma_i(N,E,\mu)=\delta_{iQ}+\delta_{i\Qb}+
\asb(\mu)\,\hat{a}^{(1)}_i(N,E,\mu)
   +{\cal O}(\asb^2(\mu))\;,
\eeq
where $\hat{a}^{(1)}_i$ vanishes unless $i$ is either $Q$, $\overline{Q}$
or $g$.
Thus, combining eq.~(\ref{eq:partN}) with eq.~(\ref{eq:integAPN})
according to eq.~(\ref{eq:sigmaN}), we obtain
\beqn
 \sigma(N,E,m) &=& 1+
 \asb(\mu)\left[ \hat{a}^{(1)}_Q(N,E,\mu)+d^{(1)}_Q(N,\mu_0,m)
 +P_{QQ}^{(0)}(N) \log\frac{\mu^2}{\mu_0^2} \right]
\nonumber \\
&+& \asb^2(\mu)\Bigg\{\sum_i \hat{a}^{(1)}_i(N,E,\mu) P^{(0)}_{iQ}\,
\log\frac{\mu^2}{\mu_0^2}
+ 2\pi\,b_0\,d^{(1)}_Q(N,\mu_0,m)\;\log\frac{\mu^2}{\mu_0^2}
\nonumber \\ &+&
 \sum_j \left[ P_{Qj}^{(0)}(N) +  P_{\Qb j}^{(0)}(N)\right] \,
 d^{(1)}_j(N,\mu_0,m)\;\log\frac{\mu^2}{\mu_0^2}
 \nonumber \\ &+&
 \sum_{k} \left[ P_{Qk}^{(0)}(N) + P_{\Qb k}^{(0)}(N)\right]\;P_{kQ}^{(0)}(N)
 \;\frac{1}{2}\log^2\frac{\mu^2}{\mu_0^2}
 \nonumber \\ \label{eq:sigmaNall} &+&
 \left[ P_{QQ}^{(1)}(N)+ P_{\Qb Q}^{(1)}(N)\right] \,
 \log\frac{\mu^2}{\mu_0^2} + \pi\,b_0\, P_{QQ}^{(0)}(N)
 \;\log^2\frac{\mu^2}{\mu_0^2} \Bigg\}
\;.
\eeqn
The above formula should accurately describe the terms of order
$\asb L$, $\asb$, $\asb^2 L^2$ and $\asb^2 L$. Terms of order
$\asb^2$, without logarithmic enhancement, are not accurately given
by the fragmentation formalism at NLO level, and have consistently been
neglected.

The lowest order splitting functions are given by
\beqn
P^{(0)}_{QQ}(N) &=& 
  C_F\left[\frac{3}{2}+\frac{1}{N(N+1)}-2\,S_1(N)\right]\,,
\nonumber \\
P^{(0)}_{Qg}(N) &=& P^{(0)}_{\Qb g}(N) = 
         C_F\left[\frac{2+N+N^2}{N(N^2-1)}\right]\,,
\nonumber \\
P^{(0)}_{gQ}(N) &=& T_F\left[ \frac{2+N+N^2}{N(N+1)(N+2)}\right]\,,
\eeqn
where, restricting ourselves to integer values of $N$,
\beq
S_1(N)=\sum_{j=1}^N\frac{1}{j}\,.
\eeq
$P_{QQ}^{(1)}(N)$ and $P_{\Qb Q}^{(1)}(N)$ are given by
\beq\label{QQandQBARQ}
P_{QQ}^{(1)}(N) = P^{\rm NS}_{QQ}(N) + P_{q^\prime q}(N)\,,\quad
P_{\Qb Q}^{(1)}(N) = P^{\rm NS}_{\Qb Q}(N) + P_{q^\prime q}(N)\,,
\eeq
where the non-singlet components are given by
\beq
\begin{array}{ll}
P_{QQ}^{\rm NS}(N) = P^{C_F}_{QQ}(N)+ P^{C_A}_{QQ}(N)+P^{n_f}_{QQ}(N)\,,
&
P_{\Qb Q}^{\rm NS}(N) = P^{C_F}_{\Qb Q}(N)+ P^{C_A}_{\Qb Q}(N)\,,
\\[.2cm]
P^{C_F}_{QQ}(N)=C_F^2\left[P_F(N)+\Delta(N)\right]\,,
&
P^{C_A}_{QQ}(N)=\frac{1}{2} C_F C_AP_G(N)\,,
\\[.2cm]
P^{n_f}_{QQ}(N)=n_f C_F T_F P_{NF}(N)\,,
&
P^{C_F}_{\Qb Q}(N)=C_F^2 P_A(N)\,,
\\[.2cm]
P^{C_A}_{\Qb Q}(N)=-\frac{1}{2} C_F C_A P_A(N)\,,
&
\end{array}
\eeq
and
\beq
P_{q^\prime q}(N)
 = -C_F T_F \frac{8 + 44 N + 46 N^2 + 21 N^3 + 14 N^4 + 15 N^5 + 
      10 N^6 + 2 N^7}{N^3 (N+1)^3(N+2)^2 (N-1)}\,.
\eeq
$P_F(N)$, $\Delta(N)$, $P_G(N)$ and $P_{NF}(N)$ were taken
from the appendix of ref.~\cite{MeleNason} and $P_A(N)$ is given
in eq.~(5.39) of ref.~\cite{CurciFurmanskiPetronzio}.
We have obtained our explicit expression for $P_{q^\prime q}(N)$
using the equation
\beq \label{pqpq}
P_{q^\prime q}=\frac{P^{\rm S}_{QQ}-P^{\rm NS}_{QQ}
 -P^{\rm NS}_{\Qb Q}}{2n_f}\,,
\eeq
where $P^{\rm S}_{QQ}$ is the singlet component\footnote{We warn the reader
that, sometimes, in the literature, the notation $P^{\rm S}$ is used for
the ``sea'' component, and $P_{QQ}$ is used for the singlet one.
Here we use $P_{QQ}$ for the full $QQ$ splitting function.}, calculated
in ref.~\cite{FurmanskiPetronzio}.
Equation~(\ref{pqpq}) is easily seen to follow from
eqs.~(\ref{QQandQBARQ}) and from eqs.~(2.42) of ref.~\cite{NasonWebber}.

The expressions for
$\hat{a}_{Q}^{(1)}$ and $d^{(1)}_{Q}$ are respectively given in eq.~(A.12)
and (A.13)  of ref.~\cite{MeleNason}.
The coefficient $\hat{a}_{g}^{(1)}$ can be obtained by performing the
Mellin transform of the expression $c_{\rm T,g}+c_{\rm L,g}$,
where $c_{\rm T,g}$ and $c_{\rm L,g}$ are given in eq.~(2.16) of
ref.~\cite{NasonWebber}. Thus
\beqn
\hat{a}_{g}^{(1)}(N,E,\mu) &=& C_F \Biggl\{
\frac{2(2+N+N^2)}{N(N^2-1)} \log\frac{E^2}{\mu^2} 
+ 4 \left[ -\frac{2}{(N-1)^2} + 
\frac{2}{N^2} -\frac{1}{(N+1)^2} \right] 
\nonumber\\
&& - 2 \left[ \frac{2}{N-1} S_1(N-1) -\frac{2}{N} S_1(N) + 
\frac{1}{N+1} S_1(N+1)  \right]
\Biggr\}
\nonumber \\
d^{(1)}_{g}(N,\mu_0,m)&=&P^{(0)}_{gQ}(N)\log\frac{\mu_0^2}{m^2}\,.
\eeqn
In order to make a more detailed comparison with our fixed order calculation, 
we separate the ${\cal O}(\asb^2)$ contributions to $\sigma(N,E,m)$
according to their colour factors. Choosing for simplicity
$\mu=E$ and $\mu_0=m$, and using the notation
\beq
\hat{a}^{(1)}_{Q/g}(N)=\hat{a}^{(1)}_{Q/g}(N,E,\mu)\vert_{\mu=E}\,,\quad
d^{(1)}_{Q/g}(N)=d^{(1)}_{Q/g}(N,\mu_0,m)\vert_{\mu_0=m}\,,
\eeq
we write
\newcommand\tsq{\log^2\frac{E^2}{m^2}}
\renewcommand\t{\log\frac{E^2}{m^2}}
\beqn
\sigma(N,E,m) &=& 1+ \asb(E)\, A(N,E,m) + \asb^2(E)\, B(N,E,m)
\nonumber \\
A(N,E,m)&=& \hat{a}^{(1)}_Q(N)+d^{(1)}_Q(N) + P_{QQ}^{(0)}(N) \t
\nonumber \\
B(N,E,m)&=&B_{C_F}(N,E,m)+B_{C_A}(N,E,m)
                     +B_{n_f}(N,E,m)+B_{T_F}(N,E,m)
\nonumber \\
B_{C_F}(N,E,m)&=&\left\{ P^{(0)}_{QQ}(N) \left[ d^{(1)}_{Q}(N) + 
               \hat{a}^{(1)}_{Q}(N)\right] +
              P^{C_F}_{QQ}(N)+ P^{C_F}_{\Qb Q}(N) \right\} \t
\nonumber \\ &&+ \frac{1}{2} \left[P^{(0)}_{QQ}(N) \right]^2 \tsq
\nonumber \\
B_{C_A}(N,E,m)&=&\left[P^{C_A}_{QQ}(N) + P^{C_A}_{\Qb Q}(N)+
         \frac{11}{6}C_A\,
              d^{(1)}_{Q}(N)\right] \t
\nonumber \\ &&+ \frac{11}{12} C_A \, P^{(0)}_{QQ}(N) \tsq
\nonumber \\
B_{n_f}(N,E,m)&=&\left[ P^{n_f}_{QQ}(N) -\frac{2}{3} n_f T_F
              d^{(1)}_{Q}(N) \right] \t
\nonumber \\      &&-\frac{1}{3}n_f T_F P^{(0)}_{QQ}(N) \tsq
\nonumber \\
B_{T_F}(N,E,m)&=&\left[\hat{a}_g^{(1)}(N)\, P_{gQ}^{(0)}
        +2\,P_{Qg}^{(0)}(N)\,d^{(1)}_g(N) + 2\, P_{q^\prime q}(N)\right]\t
\nonumber \\ && +P_{Qg}^{(0)}(N)\, P_{gQ}^{(0)}(N)\tsq\;,
\label{eq:colcoef}
\eeqn
where the subscripts $C_F$, $C_A$, $n_f$ and $T_F$ denote the
$C_F^2$, $C_F C_A$, $n_f C_F T_F\,$ and $C_F T_F$ colour components.

\begin{figure}[htb]
\centerline{\epsfig{figure=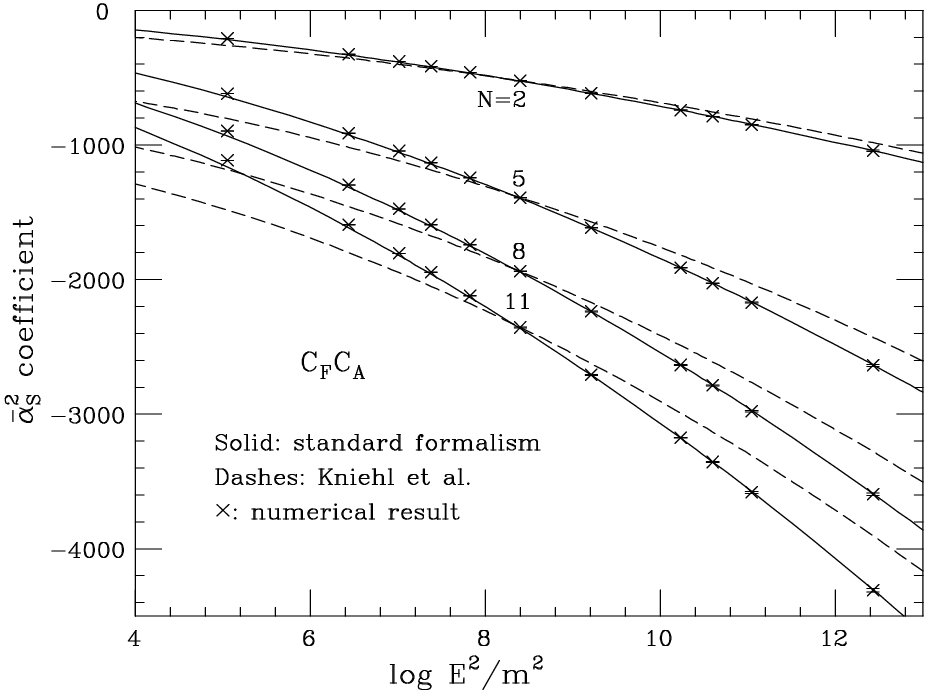,width=0.7\textwidth,clip=}}
\ccaption{}{ \label{fig:ca}
$C_FC_A$ component of the $\asb^2$ coefficient in $\sigma(N,E,m)$,
as a function of $\log E^2/m^2$, for $N=$2, 5, 8 and 11.}
\end{figure}

\begin{figure}[htb]
\centerline{\epsfig{figure=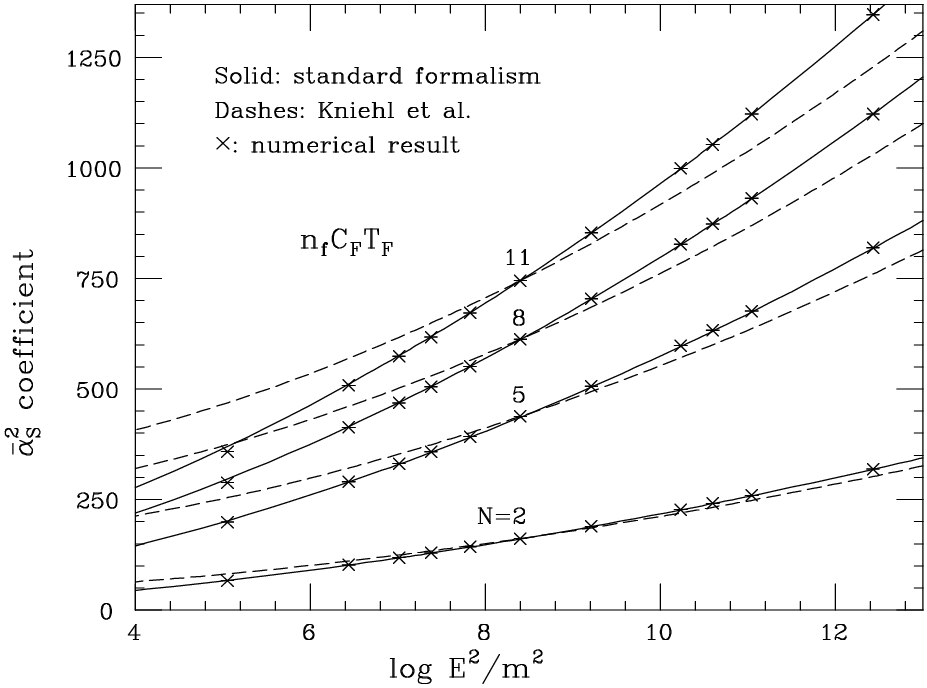,width=0.7\textwidth,clip=}}
\ccaption{}{ \label{fig:nf}
Same as in Fig.~\protect{\ref{fig:ca}}, for the $ n_f C_F T_F$ component.}
\end{figure}

\begin{figure}[htb]
\centerline{\epsfig{figure=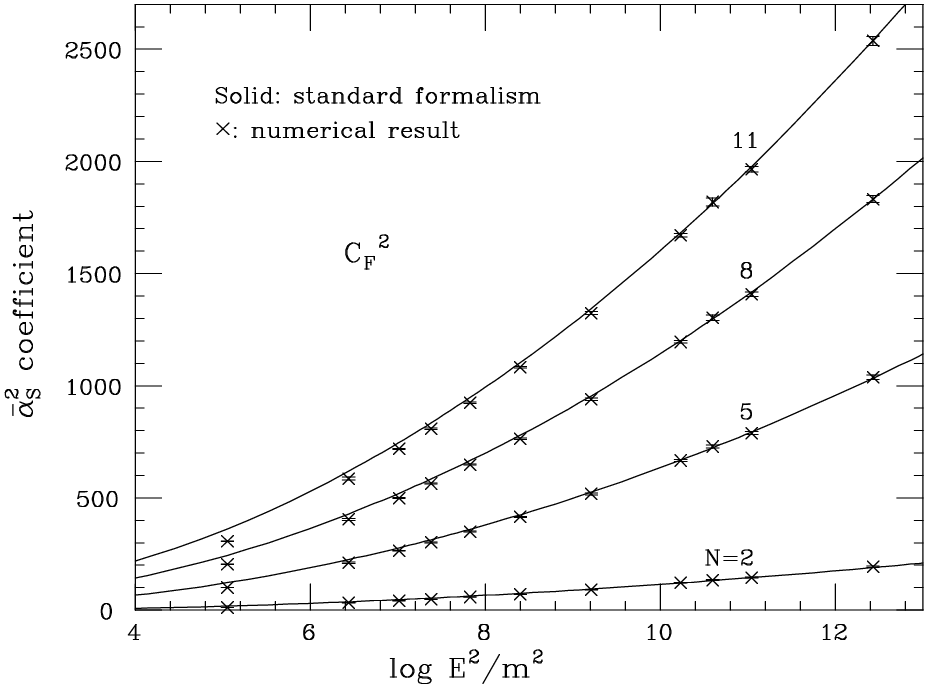,width=0.7\textwidth,clip=}}
\ccaption{}{ \label{fig:cf}
Same as in Fig.~\protect{\ref{fig:ca}}, for the $C_F^2$ component.}
\end{figure}

\begin{figure}[htb]
\centerline{\epsfig{figure=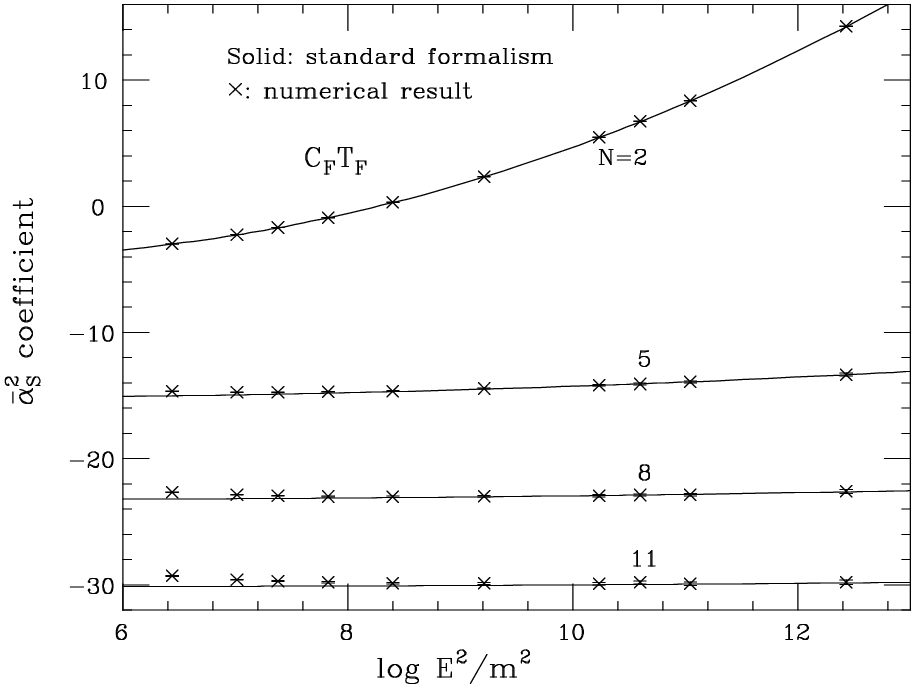,width=0.7\textwidth,clip=}}
\ccaption{}{ \label{fig:tf}
Same as in Fig.~\protect{\ref{fig:ca}}, for the $C_F T_F$ component.}
\end{figure}

The fixed order calculation of ref.~\cite{NasonOleari}
can be used to compute the cross section for the production
of a heavy quark pair plus one or two more partons, at order $\asb^2$.
We separate contributions in which four heavy quarks are present in the final
state, from those where a single $Q\Qb$ pair is present together with
one or two light partons.
These last contributions were computed only in a three-jet configuration, and
they are singular in the two-jet limit,
that is to say, when $x\to 1$.
Furthermore, the virtual corrections to the two-body process
$V\,\to\, Q+\Qb$ are not included in our calculation.
In order to remedy for these problems, we proceed as follows.
The ${\cal O}(\asb^2)$ inclusive cross section for
$V\,\to\, Q+\Qb+X$, can be written symbolically in the following form
\beq
\frac{d\s}{dx} =
   a^{(0)}\delta(1-x) +  \asb\int dY \,a^{(1)}(x,Y)
+  \asb^2 \left[\int dY\, a^{(2)}_l(x,Y)\,
+ 2 \int dY\, a^{(2)}_h(x,Y) \right]
\eeq
where $Y$ denotes all the other
kinematical variables, besides $x$, upon which the final state may depend.
We assume $\mu=E$, and
we do not indicate, for ease of notation, the dependence upon $E$ and $m$
of the various quantities. The term $a^{(2)}_l$
arises from final states with a single $Q\Qb$ pair plus at most two light
partons, while $a^{(2)}_h$ arises from final states with two $Q\Qb$
pairs. The factor of 2 in front of the $a^{(2)}_h$ contribution accounts
for the fact that we may detect either one of the two heavy quarks. 
The moments of the inclusive cross section can be written in the following way
\beqn
\s(N)=\int dx\, x^{N-1}\,\frac{d\s}{dx} \; =\; \sigma\,
+ \, \asb\int dx\,dY \,(x^{N-1}-1)\,a^{(1)}(x,Y) \phantom{aaaaaAAAAA}&&
\nonumber \\
{}+\asb^2 \left[\int dx\,dY \,(x^{N-1}-1)\,a^{(2)}_l(x,Y)\,
+ 2\int dx\,dY \,(x^{N-1}-{\scriptstyle \frac{1}{2}})\, 
a^{(2)}_h(x,Y)\right]\phantom{a}&&
\eeqn
where
\beq
\sigma =    a^{(0)} +  \asb\int dx\,dY \,a^{(1)}(x,Y) 
+  \asb^2 \left[\int dx\, dY\, a^{(2)}_l(x,Y)\,
+\int dx\, dY\, a^{(2)}_h(x,Y) \right]\;.
\eeq
The expression for $\s(N)$ can  now be easily computed with our program,
since the \mbox{$(x^{N-1}-1)$} factors regularize
the singularities in the two-jet limit,
and suppress the two-body $V\,\to\, Q+\Qb$
virtual terms.
Furthermore, in the massless limit 
\beq
\sigma = 1 + 2\,\asb(E) + c\,\asb^2(E)\,
       + {\cal O}\left(\frac{m^2}{E^2}\right) + {\cal O}(\asb^3)
\eeq
where $c$ is a constant. In fact,
the ${\cal O}(\asb^2)$ term does not contain any large logarithm,
as long as $\asb$ is the coupling with $n_f$ flavours, including the heavy
one\footnote{If instead the cross section formulae are expressed in terms
of $\asb^{(n_f-1)}$ we have\newline
$\sigma = 1 + 2\,\asb^{(n_f-1)}(E) + \frac{4}{3} T_F 
\log\frac{E^2}{m^2}\asb^2 + {\cal O}(\asb^2)$.}.
Reintroducing the energy and mass dependence, we have
\beqn
\s(N,E,m) &=& 1 +
 \asb(E)\, L(N,E,m)+\asb^2(E)\, M(N,E,m) \nonumber \\
    &&\quad\quad\quad\quad +\;
     c\,\asb^2(E)+ {\cal O}\left(\frac{m^2}{E^2}\right) +{\cal O}(\asb^3)
\nonumber \\
L(N,E,m)&=& 2 + \int dx\, dY \(x^{N-1}-1\)a^{(1)}(x,Y)
 \nonumber \\
M(N,E,m)&=& \int dx\,dY\, \(x^{N-1}-1\)
  a^{(2)}_l(x,Y) \nonumber \\
  && {}+ 2\,\int dx\, dY\, \(x^{N-1}-{\scriptstyle\frac{1}{2}}\)
  a^{(2)}_h(x,Y)\;.
\eeqn
We have calculated $L(N,E,m)$ and $M(N,E,m)$ numerically,
using $E = 100$ GeV and $m=$ 8, 4, 3, 2.5, 2, 1.5, 1, 0.6,
0.5, 0.4, 0.2~GeV,
for a vector current coupled to the heavy quark.
We expect that, for small masses, $A(N,E,m)$ should coincide with
$L(N,E,m)$, and $M(N,E,m)$ should differ from $B(N,E,m)$ by a
mass and energy independent quantity, since such term is actually
beyond the next-to-leading logarithmic approximation. We find
very good agreement between $A(N,E,m)$ and $L(N,E,m)$.
We present the results for $M(N,E,m)$ separated into the different
colour components
\beq
M(N,E,m)=M_{C_F}(N,E,m)+M_{C_A}(N,E,m)
                     +M_{n_f}(N,E,m)+M_{T_F}(N,E,m)\;.
\eeq
In Fig.~\ref{fig:ca} we have plotted our results for $M_{C_A}$ (crosses
with error bars) and for $B_{C_A}$ (solid lines).
An arbitrary ($N$-dependent) constant has been added to the
curves for $B_{C_A}$, in order to make them coincide with the numerical result
for $m/E=0.015$.
We find, as the mass gets smaller,
satisfactory agreement for all moments.
Notice that, as intuitive reasoning would suggest, for higher moments
we need smaller masses to approach the massless limit.
In Figs.~\ref{fig:nf}, \ref{fig:cf} and~\ref{fig:tf} we report the analogous
results for the remaining colour combinations.
Again, we find satisfactory agreement.

If one follows the procedure proposed by Kniehl et al.,
eqs.~(\ref{eq:colcoef}) are modified in the $C_F C_A$ and in the $n_f C_F T_F$
coefficients. More specifically, the terms proportional to $d^{(1)}_Q$
all disappear from the expressions of the  $C_F C_A$ and of the $n_F C_F T_F$
coefficients. In fact, by inspecting formulae~(\ref{eq:sigmahatprime})
and~(\ref{eq:iniprime}), and the derivation of eq.~(\ref{eq:sigmaNall}),
we see that the only relevant difference between the
two approaches is that the term $\asb(\mu_0)\,d^{(1)}_Q$
is replaced by $\asb(\mu)\,d^{(1)}_Q$, which, using the renormalization
group equation, amounts to a difference of
$-2\pi\, b_0\, \asb^2 \,d^{(1)}_Q\,\log\mu^2/\mu_0^2 $, precisely what is
needed to cancel the term of the same form appearing in
eq.~(\ref{eq:sigmaNall}). The modified result of Kniehl et al.\ is also shown
in Figs.~\ref{fig:ca} and~\ref{fig:nf} (dashed lines).
It is quite clear that their approach does not work.

\section{Conclusions}
In the present work, we have verified at order $\asb^2$ the NLO
fragmentation function approach to the computation of the heavy quark
fragmentation function given in ref.~\cite{MeleNason}.
Besides excluding an alternative that has been proposed in the
literature~\cite{Spira}, 
we have given a verification of several ingredients that go into
the fragmentation function formalism, such as the initial conditions,
computed in ref.~\cite{MeleNason}, the NLO splitting functions in the time-like
region~\cite{CurciFurmanskiPetronzio}, and finally we have also
performed a further test of the validity of the fixed order calculation
of ref.~\cite{NasonOleari}.


\end{document}